# On the Hamiltonian nature of semiclassical equations of motion in the presence of an electromagnetic field and Berry curvature

K.Yu. Bliokh[a,b,c*]

[a] *Institute of Radio Astronomy, 4 Krasnoznamennaya st., Kharkov, 61002, Ukraine*
[b] *A.Ya. Usikov Institute of Radiophysics and Electronics, 12 Akademika Proskury st., Kharkov, 61085, Ukraine*
[c] *Department of Physics, Bar-Ilan University, Ramat Gan, 52900, Israel*

We consider the semiclassical equations of motion of a particle when both an external electromagnetic field and the Berry gauge field in the momentum space are present. It is shown that these equations are Hamiltonian and relations between the canonical and covariant variables are determined through a consistent account of all components of the Berry connection. The Jacobian of the canonical-to-covariant-variables transformation describes the nonconservation of the 'naive' phase space volume in the covariant coordinates (D. Xiao, J. Shi, and Q. Niu, Phys. Rev. Lett **95**, 137204 (2005)).

In recent years, a great attention has been focused on the re-examination of semiclassical equations of motion (EOM) for various quantum particles (both free and in condensed matter) in external fields. The main new point is that that in multi-level system the Berry gauge field (Berry curvature) contributes to EOM as a real external physical field, which influences the evolution of the particle [1–4]. This has led to the discovering and explanation of various physical effects such as the anomalous Hall effect, intrinsic spin Hall effect, optical Magnus effect, quantum spin pumping etc. [4–8].

Lately, it was noted that the simultaneous introduction of the Berry gauge field in the momentum space and of the magnetic field meets with an additional difficulties connected to the non-commutativity of the two fields [8,9]. By analyzing semiclassical dynamics of a Bloch electron in an external magnetic field with the Berry curvature taken into account, Xiao, Shi, and Niu [9] came to a conclusion that it is described by non-canonical equations with the phase space volume non-conserved (violation of the Liouville's theorem), and that leads to a number of observable physical phenomena. In this Letter we show that the particle's EOM represent, in actual fact, Hamiltonian equations in covariant coordinates, whose connection with canonical coordinates is determined by *all* components of the Berry gauge potential (Berry connection). Such a Hamiltonian approach has been realized in [8] for a description of Dirac electron and has led to similar to [9] EOM, commutation relations and observable phenomena.

Let us consider adiabatic evolution of a charged particle corresponding to a certain (non-degenerated) energy level of a multilevel system. If the initial Hamiltonian of the system was non-diagonal and generated the Berry connection in the momentum space, then the semiclassical EOM of the particle in an external electromagnetic field have the form [3–5,8]

$$\dot{r}^\alpha = \partial_{p_\alpha} \mathrm{H} - \hbar \mathcal{F}^{\alpha\beta} \dot{p}_\beta, \quad \dot{p}_\alpha = -\partial_{r^\alpha} \mathrm{H} + eF_{\alpha\beta} \dot{r}^\beta. \tag{1}$$

We represent Eqs. (1) in 4D form in order to stress their symmetry and take the possibility of relativistic applications into account. In Eqs. (1) $c=1$, $e$ is the particle's charge, $r^\alpha = (t, \mathbf{r})$ and $p_\alpha = (-E, \mathbf{p})$ are classical 4-coordinates and conjugate 4-momentum of the particle, $\mathrm{H}(r^\alpha, p_\alpha) = H(r^\alpha, \mathbf{p}) + p_0 = 0$ is the Hamiltonian modified for 4D form of the equations [4,8,10], $F_{\alpha\beta}(r^\alpha) = \partial_{r^\alpha} \wedge A_\beta$ is the electromagnetic field tensor ($A_\alpha = (-\varphi, \mathbf{A})$ is the electromagnetic potential), and $\mathcal{F}^{\alpha\beta}(p_\alpha) = \partial_{p_\alpha} \wedge \mathcal{A}_\beta$ is the Berry gauge field tensor ($\mathcal{A}_{p_\alpha} = (0, \mathcal{A}(\mathbf{p}))$ is the Berry gauge potential in momentum space). Only space components of $\mathcal{F}^{\alpha\beta}$ are non-zero; thus it can be represented in the form of axial vector $\mathcal{B} = \mathrm{curl}\,\mathcal{A}$, "conjugated" to the magnetic field $\mathbf{B} = \mathrm{curl}\,\mathbf{A}$.

---

[*] E-mail: k_bliokh@mail.ru

To bring Eqs. (1) to canonical Hamiltonian form, note that variables $r^\alpha$ and $p_\alpha$ are covariant ones and are connected to canonical (generalized) variables, $R^\alpha$ and $P_\alpha$, by means of gauge potentials: electromagnetic and Berry's. The initial Hamiltonian (which was written as a function of canonical variables) in the absence of the magnetic field contained the non-diagonality dependent on momentum $\mathbf{P}$, that led to the Berry potential in the momentum space only. However, after the introduction of the magnetic field, $\mathbf{P} \to \mathbf{P} - e\mathbf{A}(R^\alpha)$, the non-diagonal part of the Hamiltonian becomes dependent on coordinates $R^\alpha$ as well. Hence, this leads to the appearance of the Berry gauge potential on the coordinate space as well. Since the Hamiltonian depends on the combination $\mathbf{P} - e\mathbf{A}(R^\alpha)$ only, the space components of the Berry connection can be easily expressed by means of the momentum ones: $\mathcal{A}_{R^\alpha} = -e\mathcal{A}_{p_\beta} \partial_{R^\alpha} A_\beta$ ($\mathcal{A}_\alpha \equiv \mathcal{A}_{p_\alpha}$). With space components of the Berry connection taken into account, the relation between canonical and covariant variables takes the form

$$r^\alpha = R^\alpha + \hbar\mathcal{A}_{p_\alpha}(p_\alpha) = R^\alpha + \hbar\mathcal{A}_{p_\alpha}(P_\alpha - eA_\alpha(R^\alpha)),$$
$$p_\alpha = P_\alpha - eA_\alpha(r^\alpha) - \hbar\mathcal{A}_{R^\alpha}(r^\alpha, p_\alpha) = P_\alpha - eA_\alpha(R^\alpha) + e\hbar F_{\alpha\beta}(R^\alpha)\mathcal{A}_{p_\beta}(P_\alpha - eA_\alpha(R^\alpha)). \quad (2)$$

All calculations are carried out with an accuracy of $\hbar$, with which semiclassical Eqs. (1) are written. By using Eqs. (2), one can show that Eqs. (1) indeed become Hamiltonian: $\dot{R}^\alpha = \partial_{P_\alpha} H$, $\dot{P}_\alpha = -\partial_{R^\alpha} H$. Besides, using the commutators for quantum operators corresponding to the canonical variables, $[\hat{R}^\alpha, \hat{R}^\beta] = 0$, $[\hat{P}_\alpha, \hat{P}_\beta] = 0$, $[\hat{R}^\alpha, \hat{P}_\beta] = i\hbar\delta^\alpha_\beta$, one can calculate commutators of operators of covariant variables (2) [8]:

$$-i\hbar^{-1}[\hat{r}^\alpha, \hat{r}^\beta] = \hbar\mathcal{F}^{\alpha\beta}, \quad -i\hbar^{-1}[\hat{p}_\alpha, \hat{p}_\beta] = eF_{\alpha\beta} + e^2\hbar F_{\alpha\gamma}F_{\beta\delta}\mathcal{F}^{\gamma\delta},$$
$$-i\hbar^{-1}[\hat{r}^\alpha, \hat{p}_\beta] = \delta^\alpha_\beta + e\hbar\mathcal{F}^{\alpha\gamma}F_{\beta\gamma}. \quad (3)$$

With used accuracy of $\hbar$, space components of commutators (3) coincide with the commutators derived in 2D case in [9] and in more general form in [11]. These commutators can be deduced directly from fundamental physical equations [8].

Obviously, the phase space volume is conserved in canonical variables and the Liouville's theorem holds true: $\Delta V = \Delta R^\alpha \Delta P_\alpha = \mathrm{const}$. Jacobian of the transformation of the canonical variables into the covariant ones, $D = \det(\partial Z^\alpha / \partial z^\beta) = \det(\partial Z_i / \partial z_j)$, (where $Z^\alpha \equiv (R^\alpha, P_\alpha)$, $z^\alpha \equiv (r^\alpha, p_\alpha)$, $\mathbf{Z} \equiv (\mathbf{R}, \mathbf{P})$, and $\mathbf{z} \equiv (\mathbf{r}, \mathbf{p})$), determines variations of the 'naive' phase space volume in the covariant variables: $\Delta\tilde{V} = \Delta r^\alpha \Delta p_\alpha = D^{-1}\Delta V \neq \mathrm{const}$. According to [9,11], $D = 1 - e\hbar F_{\alpha\beta}\mathcal{F}^{\alpha\beta}/2 = 1 - e\hbar\mathbf{B}\mathcal{B}$.

To conclude, we have shown that semiclassical equations of motion of a particle in the presence of an electromagnetic field and Berry curvature are Hamiltonian indeed, and the relation between canonical and covariant variables is determined by the consistent account of all components of the Berry connection. An alternative way of Hamiltonian description of Eqs. (1) is the introduction of non-canonical symplectic structure (Poisson brackets) on the phase space of covariant variables [1,9,11]. Such a symplectic structure enables one to define the true phase space volume, which satisfies the Liouville's theorem. Evidently, different approaches are equivalent to each other and represent only different formalisms.

The work was supported in part by INTAS (Grant No. 03-55-1921).